\documentclass[	
						superscriptaddress,%
						twocolumn,%
						a4paper,
						showkeys,
					]{revtex4}

\usepackage[frenchb,english]{babel}
\usepackage[T1]{fontenc}
\usepackage{upgreek}
\usepackage{textcomp} 				
\usepackage[dvips]{graphicx}		
\usepackage{color}					
\usepackage[nice]{nicefrac} 		
\usepackage{amssymb,amsmath}		
\usepackage{array}					
\usepackage{ulem}

\newcommand{\etal}{\textit{et al.}}

\newcommand{\modif}[2]{#2}
\definecolor{vert}{rgb}{0.5,0.758,0.5}
\definecolor{bleufonce}{rgb}{0,0,0.516}
\definecolor{orange}{rgb}{1,0.516,0}

\begin{document}

\title{Multiple high-pressure phase transitions in BiFeO$_3$}
\date{\today}
\author{Mael Guennou}
\affiliation{Laboratoire des Mat\'eriaux et du G\'enie Physique, CNRS, Grenoble Institute of Technology, MINATEC,
3 parvis Louis N\'eel, 38016 Grenoble, France}
\email{mael.guennou@supaero.org}
\author{Pierre Bouvier}
\affiliation{Laboratoire des Mat\'eriaux et du G\'enie Physique, CNRS, Grenoble Institute of Technology, MINATEC,
3 parvis Louis N\'eel, 38016 Grenoble, France}
\affiliation{European Synchrotron Radiation Facility (ESRF), BP 220, 6 Rue Jules Horowitz, 38043 Grenoble Cedex, France}
\email{pierre.bouvier@grenoble-inp.fr}
\author{Grace S. Chen}
\affiliation{Laboratoire des Mat\'eriaux et du G\'enie Physique, CNRS, Grenoble Institute of Technology, MINATEC,
3 parvis Louis N\'eel, 38016 Grenoble, France}
\author{Brahim Dkhil}
\affiliation{Laboratoire Structures, Propri\'et\'es et Mod\'elisation des Solides, Ecole Centrale Paris, 92290 Ch\^atenay-Malabry, France}
\author{Rapha\"el Haumont}
\affiliation{Laboratoire de Physico-Chimie de l'\'Etat Solide, ICMMO, CNRS, Universit\'e Paris XI, 91405 Orsay, France}
\author{Gaston Garbarino}
\affiliation{European Synchrotron Radiation Facility (ESRF), BP 220, 6 Rue Jules Horowitz, 38043 Grenoble Cedex, France}
\author{Jens Kreisel}
\affiliation{Laboratoire des Mat\'eriaux et du G\'enie Physique, CNRS, Grenoble Institute of Technology, MINATEC,
3 parvis Louis N\'eel, 38016 Grenoble, France}

\begin{abstract}
We investigate the high-pressure phase transitions in BiFeO$_3$ by single crystal and powder x-ray diffraction, as well as single crystal Raman spectroscopy. Six phase transitions are reported in the 0--60~GPa range. At low pressures, up to 15 GPa, 4 transitions are evidenced at 4, 5, 7 and 11~GPa. In this range, the crystals display large unit cells and complex domain structures, which suggests a competition between complex tilt systems and possibly off-center cation displacements. The non polar $Pnma$ phase remains stable over a large pressure range between 11 and 38~GPa, where the distortion (tilt angles) changes only little with pressure. The two high-pressure phase transitions at 38 and 48~GPa are marked by the occurence of larger unit cells and an increase of the distorsion away from the cubic parent perovskite cell. We find no evidence for a cubic phase at high pressure, nor indications that the structure tends to become cubic. The previously reported insulator-to-metal transition at 50~GPa appears to be symmetry breaking. 
\end{abstract}


\maketitle


\section{Introduction}

Magnetoelectric multiferroics currently attract a great deal of interest, both because of intruiging coupling mechanisms between magnetism, ferroelectricity and ferroelasticity, and because of the potential for new types of magnetoelectric devices \cite{Ramesh2007,Cheong2007,Fiebig2005,Bea2008,Kreisel2009}. Among multiferroics, Bismuth ferrite BiFeO$_3$ (BFO) is commonly considered as a model system \cite{Catalan2009}, and is perhaps the only material that is both magnetic and ferroelectric with a strong electric polarization at ambient conditions. Even though BFO attracts an important attention, the crystal structures of BFO as a function of temperature and pressure are still debated in the literature \cite{Catalan2009}.

\begin{figure}[ht]
\begin{center}
\includegraphics[width=0.4\textwidth]{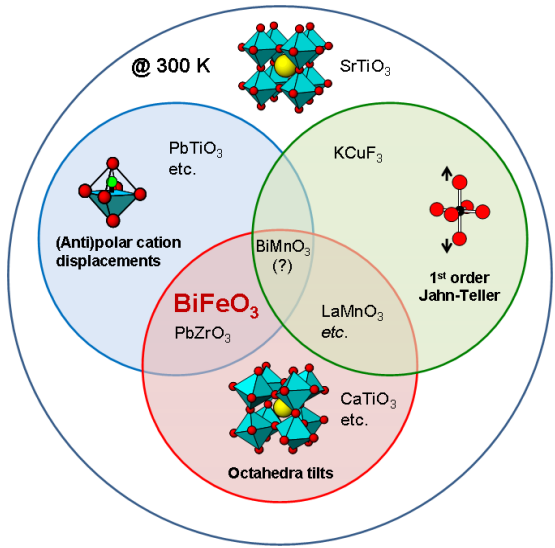}
\caption{Schematic illustration of distortions and their combination in $AB$X$_3$ perovskites. Each of the three inner circles represents one of the three main collective structural distortions in perovskites: \modif{}{(anti)polar} cation displacements (blue), octahedra tilts (red) and \modif{}{first-order} Jahn-Teller (green) distortions, each illustrated by a schematic drawing and a material that adopts the distortion at ambient conditions. Materials outside the circles adopt the ideal cubic perovskite structure exemplified by SrTiO$_3$. BiFeO$_3$ is among the very few materials that adopts at 300~K both strong \modif{}{ferroelectric} cation displacements and strong octahedra tilts.}
\label{fig:instabilities}
\end{center}
\end{figure}

Besides being a model multiferroic, BFO is also one of the very few perovskites that present both octahedra tilts \cite{Glazer1972,Mitchell2002,Glazer2011} and strong ferroelectric cation displacements, thus two structural instabilities, at room temperature. Ab-initio calculations suggest that in a number of perovskites both instabilities are present and compete \cite{Ghosez1999,Junquera2006}. It is however important to remind that one instability usually prevails so that the vast majority of perovskites presents either tilts (SrTiO$_3$, LaAlO$_3$ etc.) or ferroelectric cation displacements (BaTiO$_3$, PbTiO$_3$ etc.) at a given temperature or pressure \cite{Mitchell2002}, as schematized in Figure \ref{fig:instabilities}. For some perovskites a crossover from one distortion to the other can occur as a function of an external parameter, PbTiO$_3$ under high pressure is such an example \cite{Janolin2008}. BiFeO$_3$ is thus a rather exceptional case and turns out to be a model system for studying the competition between tilts and polar instabilities in perovskites.

General rules for predicting phase transitions in perovskites are of longstanding interest. As a fruit of past studies it is now generally accepted that increasing temperature reduces both tilt instabilities (the tilt angle decreases with increasing temperature) and polar instabilities (decrease of the polar cation displacement) \cite{Mitchell2002,Scott1969,Fleury1968}. As a consequence, increasing temperature drives perovskites towards the parent cubic perovskite, although the ideal cubic structure might be in some cases beyond the decomposition or melting point.
The effect of high-pressure on perovskites is more complex. Concerning octahedra tilts, it was originally suggested by Samara \etal{} \cite{Samara1975} that the phase transition temperatures $T_c$ of zone-boundary transitions in perovskites should always increase with pressure: $\mathrm dT_c/\mathrm dP>0$, i.e. the antiferro-distortive tilt angle increases with increasing pressure. However, experiments on LaAlO$_3$ \cite{Bouvier2002} and later on other perovskites \cite{Angel2005,Zhao2006,Zhao2004,Tohei2005} revealed that some perovskites decrease their tilt angle and undergo phase transitions to higher-symmetry phases on increasing pressure. As a consequence, Samara's rule has been extended by Angel \etal{} to a new general rule \cite{Angel2005} which predicts the behaviour of octahedra tilts by taking into account the compressibilities of the different polyhedra, with so far no known exception.

As to polar instabilities, the pioneering work by Samara \etal{} \cite{Samara1975} describes that pressure reduces ferroelectricity in $AB$O$_3$ perovskites and even annihilates it for a critical pressure $P_c$ at which the crystal structure becomes cubic. Early papers on BaTiO$_3$ \cite{Venkateswaran1998,Sood1995}, KNbO$_3$ \cite{Gourdain2002,Pruzan2007}, PbTiO$_3$ \cite{Cerdeira1975,Sanjurjo1983} amply confirmed this view and let to the widely accepted conclusion that polar perovskites adopt for $P > P_c$ a cubic $Pm\overline 3m$ crystal structure. It was thus unexpected to find out in recent years that relaxor ferroelectrics do not stay/become cubic at high-pressure (e.g. Ref. \cite{Kreisel2002,Kreisel2003,Chaabane2003,Janolin2006}) and that ferroelectric PZT displays a wealth of phase transitions under high-pressure without becoming cubic either (e.g. Ref. \cite{Sani2004,Rouquette2005,Rouquette2004}). However, the most striking observation was that the model ferroelectric and simple perovskite PbTiO$_3$ first reduces its ferroelectricity and becomes cubic but then, unlike commonly thought, becomes again non-cubic with tetragonal phases \cite{Janolin2008,Kornev2007,Kornev2005} through mechanisms involving oxygen octahedra tilting \cite{Janolin2008} and a theoretically predicted re-entrance of ferroelectricity \cite{Kornev2007,Kornev2005,Bousquet2006}.
 
The aim of our study is to go beyond the above discussed perovskites with one instability only, by investigating the consequences of the competition of octahedra tilts and cation displacements on the crystal structure of BiFeO$_3$, which is also magnetic. We focus on the effect of pressure, a parameter which allows driving to a very large extend structural instabilities, and thus their competition, thanks to otherwise unachievable large changes in cell volume and bond lengths. Despite literature reports on the effect of high-pressure on BFO, the actual symmetries in the low-pressure regime remain controversial and the very-high pressure regime has not been properly investigated (see ref. \cite{Haumont2006,Haumont2009,Guennou2011,Gavriliuk2005,Gavriliuk2006,Gavriliuk2007,Belik2009,Yang2009,Zhu2010} and following section). Here, we investigate BFO by using Raman spectroscopy and synchrotron X-ray diffraction on both powder and single crystals up to the very high-pressure of 55 GPa, a regime only rarely explored in functional oxides beside some notable exceptions \cite{Janolin2008,Guennou2010,Guennou2010a,Guennou2011}.
 
We show that BFO undergoes a surprising richness of six phase transitions up to 55 GPa. Further to this, we report for BFO unusual large and puzzling unit cells under high pressure, which are unexpected for simple $AB$O$_3$ perovskites. The results are presented in light of the current literature on BFO and also in relation to other Bi-based perovskites.


\section{Preliminary remarks}

\begin{figure}[tb]
\begin{center}
\includegraphics[width=0.48\textwidth]{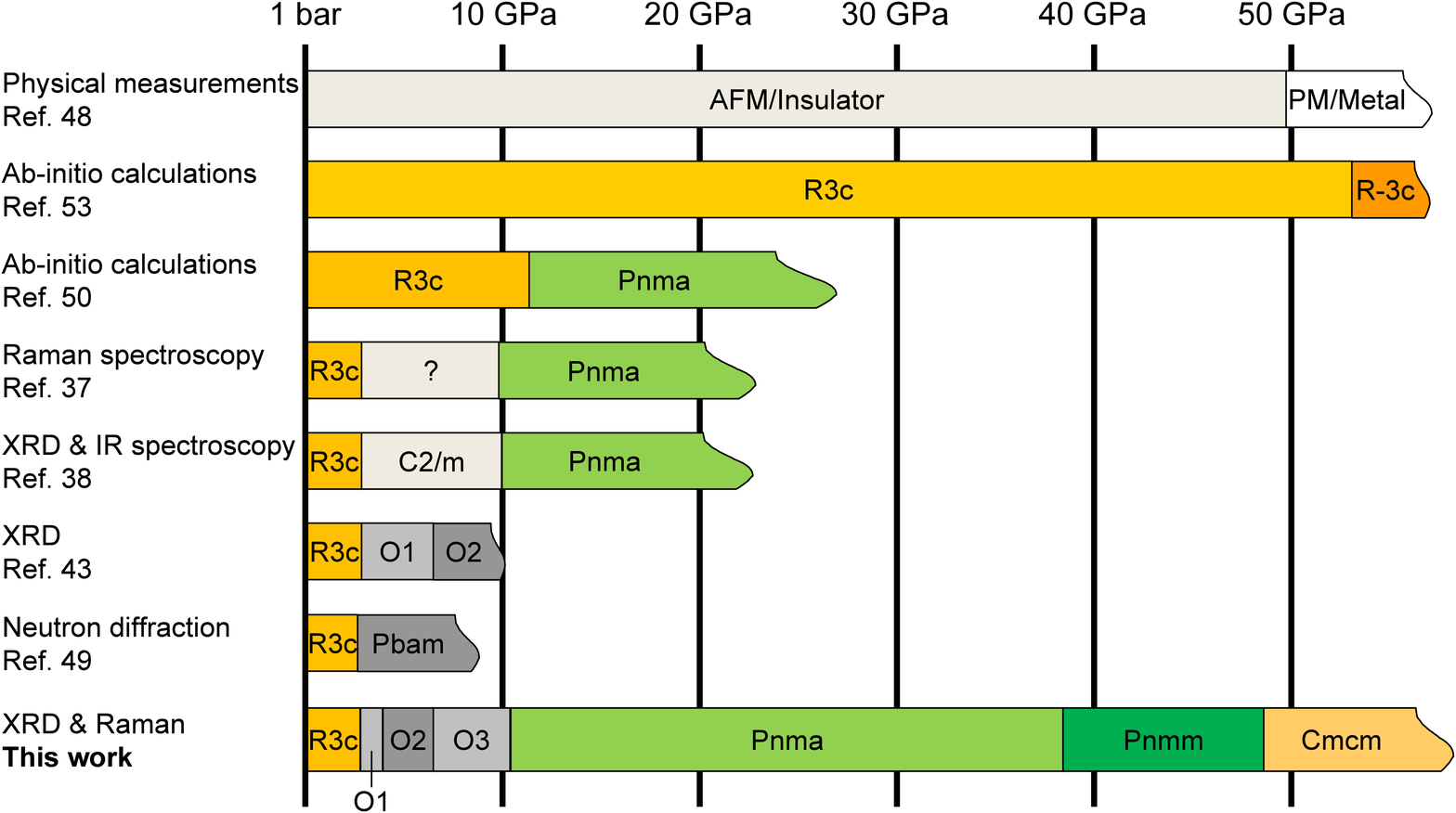}
\caption{Schematic summary of the high-pressure investigations of BFO.}
\label{fig:literature}
\end{center}
\end{figure}

The structure of BFO at high pressure has been investigated in the past in several studies, as schematized in figure \ref{fig:literature}. Powder x-ray diffraction under high-pressure has been performed over the past few years by several groups with various outcomes. Gavriliuk \etal{} \cite{Gavriliuk2007,Gavriliuk2008} and Zhu \etal{} \cite{Zhu2010} have reported no structural phase transition in this pressure range. Haumont \etal{} \cite{Haumont2009} have identified two phase transitions at 3.5 and 10~GPa with a proposed phase sequence $R3c\longrightarrow C2/m\longrightarrow Pnma$. Belik \etal{} \cite{Belik2009} have later confirmed the phase transition at 4 GPa and identified an additional transition at 7 GPa. They have identified the two intermediate phases as orthorhombic instead of the monoclinic $C2/m$, with a transition sequence $R3c\longrightarrow\mathrm{Ortho\ I}\longrightarrow\mathrm{Ortho\ II}\longrightarrow Pnma$, and also pointed out reversibility issues. In a previous work, we have confirmed these transition pressures and emphasized the importance of good hydrostatic conditions \cite{Guennou2011}. \modif{}{Very recently, Kozlenko \etal{} \cite{Kozlenko2011} have reported a powder neutron diffraction study and identified a high-pressure orthorhombic phase with space group $Pbam$ between 3 and 8.6~GPa, distinct from the phases previously seen.} In parallel, Raman spectroscopy has also been used to investigate phase transitions \cite{Haumont2006,Yang2009}. Both studies revealed two phase transitions in the low-pressure range, but at somewhat different pressures, about 3 and 9~GPa. Theoretical studies have also been devoted to the high-pressure behaviour of BFO \cite{Ravindran2006,Gonzalez-Vazquez2009,Feng2010,Dieguez2011} and will be commented on in the discussion.

Generally speaking, the determination of the structures of BFO from single crystal diffraction in diamond-anvil cells (DAC) is complicated by (i) the limited access to reciprocal space, because of the geometry of the DACs, (ii) the complex domain structure that emerges as the crystal goes through the different phase transitions (iii) the very weak intensity of oxygen-related superstructure reflections (SSR) as compared to the heavier Bi and Fe. Powder diffraction on the other hand is not limited by the DAC geometry but overlapping bands, band broadening and a stronger background hamper the observation of weak reflections and small splittings. Both techniques can be used in a complementary way, although single crystal data give in general much better results, especially at high pressure. In this paper, we report four different diffraction experiments carried out on powder and single crystal samples. The experiments are summarized in table \ref{tab:experimentlist} with the pressure ranges investigated. In the following, the diffraction results will be mostly taken from experiments 1 and 2 carried out with single crystals. Experiments 3 and 4 were carried out to check the reproducibility of the transitions; they will not be presented in detail. 


\section{Experimental details}

\begin{table}
\begin{center}
\renewcommand{\arraystretch}{1.2}
\begin{tabular}{l l l l l}
\hline\hline
Exp. 	& Sample 			& PTM 		& Beamline & $P$ range \\\hline
XRD 1	& Single crystal 	& Helium 	& ID09A 	& 0--12 GPa \\
XRD 2 & Single crystal 	& Neon 		& ID27 	& 12--55 GPa \\
XRD 3	& Single crystal 	& Helium 	& ID27 	& 30--55 GPa \\
XRD 4	& Powder 			& Helium 	& ID27 	& 0--52 GPa \\
Raman & Single crystal 	& alcohol	&			& 0--12 GPa \\
\hline\hline
\end{tabular}
\caption{List of the experiments reported in this paper with the pressure range investigated and the pressure transmitting media (PTM) used.}
\label{tab:experimentlist}
\end{center}
\end{table}

The BFO powder was prepared by conventional solid-state reaction using high-purity (better than 99.9\%) bismuth oxide Bi$_2$O$_3$ and iron oxide Fe$_2$O$_3$ as starting compounds. After mixing in stoichiometric proportions, powders were calcined at $T_f=820^\circ$C for 3~h. BFO single crystals were grown by the flux method. Details about the synthesis can be found in Ref. \cite{Haumont2008}. The single crystals used for the experiments were polished to a thickness of about 10 $\upmu$m with a lateral extension of 10 to 30~$\upmu$m. Both optical inspection under polarized light and XRD indicate that the crystals used are in a single domain state before the experiments. 

All experiments were performed in diamond-anvil cells (DAC). The diamonds have the Boehler-Almax design with cullets of 600~$\upmu$m (exp. 1 + Raman) or 250~$\upmu$m (exp. 2, 3 and 4). The pressure chamber was sealed by a stainless steel gasket. The pressure transmitting media (PTM) used are given in table \ref{tab:experimentlist}. 

X-ray diffraction experiments were performed on the ID27 and ID09A beamlines at the ESRF. At ID27, the beam was monochromatic with a wavelength of 0.3738~\AA\ selected by an iodine K-edge filter and focused to a beam size of about 3~$\upmu$m. At ID09A, the beam size is about 20~$\upmu$m and the wavelength (0.4144~\AA) was determined from the calibration using a standard silicon powder. The signal was collected in the rotating crystal geometry on a marCCD 345 (ID27) or mar 555 image plate (ID09A) with $-30^\circ\le\omega\le 30^\circ$ in 0.5 or 2$^\circ$ steps. A precise calibration of the detector parameters was performed with a reference silicon powder. The pressure was measured by the standard ruby fluorescence method \cite{Mao1986}. The diffraction patterns from single crystal measurements were indexed with a home-made program based on the Fit2D software \cite{Hammersley1996}. The refinement of the lattice constants from the peak positions was performed with the program UnitCell \cite{UnitCell}. The CrysAlis software (Oxford Diffraction--Agilent) was used for reconstruction of planes in reciprocal space. For the powder experiment, a gold powder was used as a pressure standard \cite{Dewaele2004}. The pressure calculated from the gold standard was corrected for non-hydrostatic effects by following the procedure described by Singh \cite{Singh2004}.

The Raman spectra were recorded on a Jobin-Yvon Labram spectrometer with a low frequency cutoff at 100 cm$^{-1}$. The exciting laser line was 633~nm. The laser power was kept at 10~mW on the DAC to avoid heating of the sample. The spectra are unpolarized.


\section{Results}

The analysis of the diffraction patterns reveals 6 phase transitions at 4, 5, 7, 11, 38 and 48 GPa with the phase sequence $R3c\longrightarrow\mathrm{Phase\ II}\longrightarrow\mathrm{Phase\ III}\longrightarrow\mathrm{Phase\ IV}\longrightarrow Pnma\longrightarrow\mathrm{Phase\ VI}\longrightarrow\mathrm{Phase\ VII}$. The $R3c$ and $Pnma$ phases have been reported previously and are only confirmed by our experiment. In contrast, the structure of phases II, III, IV, VI and VII have not been determined. In the following, we divide the results into two distinct pressure ranges: the low pressure region from ambient pressure up to 15 GPa and the high pressure region for the $Pnma$ phase and above. Finally we give the evolution of the lattice constants, volume and strains over the full pressure range.

\subsection{Low pressure region: 0--15 GPa}

\begin{figure}[tb]
\begin{center}
\includegraphics[width=0.48\textwidth]{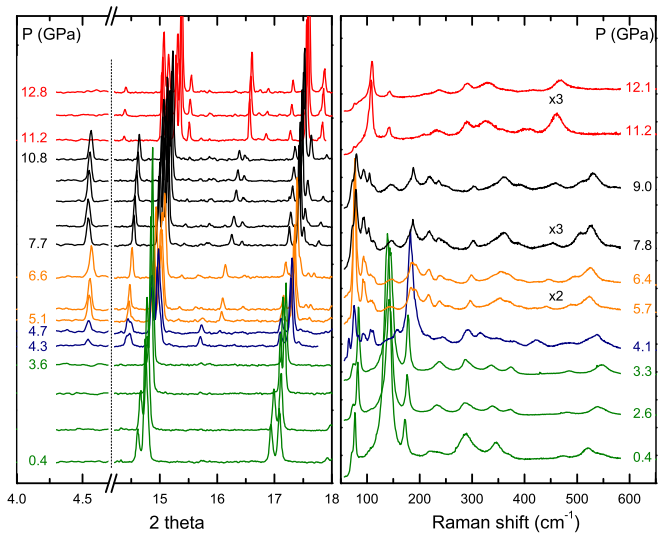}
\caption{(Left) Selected parts of the integrated diffraction patterns of the single crystal for the low-pressure experiment ($\lambda = 0.4144$~\AA) \modif{}{with phases $R3c$ (\textcolor{vert}{green}), II (\textcolor{bleufonce}{blue}), III (\textcolor{orange}{orange}), IV (\textcolor{black}{black}) and $Pnma$ (\textcolor{red}{red}).} (Right) Raman spectra up to 12.1 GPa.}
\label{fig:overviewXRD}
\end{center}
\end{figure}

An overview of the experimental evidence for the 4 phase transitions in the low pressure range is given in figure \ref{fig:overviewXRD} where we present the integrated single crystal diffraction pattern (left) and the Raman spectra (right).

The transition from the low-pressure $R3c$ phase to phase II is marked by the emergence of new superstructure reflections \modif{}{(blue in figure \ref{fig:overviewXRD})}. In addition, splittings of diffraction peaks show the emergence of a complex domain structure. A simple observation under a microscope also shows that the single domain has split into a multitude of very small domains in phase II.

At 5~GPa, we observe the disappearance of many diffraction peaks, revealing a simplification of the domain structure, and a marked change of the Raman signature, marking the transition from phase II to phase III \modif{}{(orange in figure \ref{fig:overviewXRD})}. The transition from phase III to phase IV \modif{}{(black in figure \ref{fig:overviewXRD})} at 7~GPa is not characterized by pronounced change in the diffraction pattern, in terms of emergence or disappearance of diffraction peaks. A close inspection of the Raman spectra does reveal small changes at 7~GPa, notably at low wavenumbers with the emergence of a mode at 100~cm$^{-1}$. The most clear signature of this transition are sudden shifts of diffraction peaks at low 2$\theta$ toward lower angles, which can only be associated with sudden changes in the lattice constants. This change coincides well with the transition seen at the same pressure in the powder diffraction experiment by Belik \etal{} \cite{Belik2009}. At 11~GPa, we observe the transition to the $Pnma$ phase by changes in both the diffraction patterns and the Raman signature \modif{}{(red in figure \ref{fig:overviewXRD})}.

\begin{figure}[tb]
\begin{center}
\includegraphics[width=0.47\textwidth]{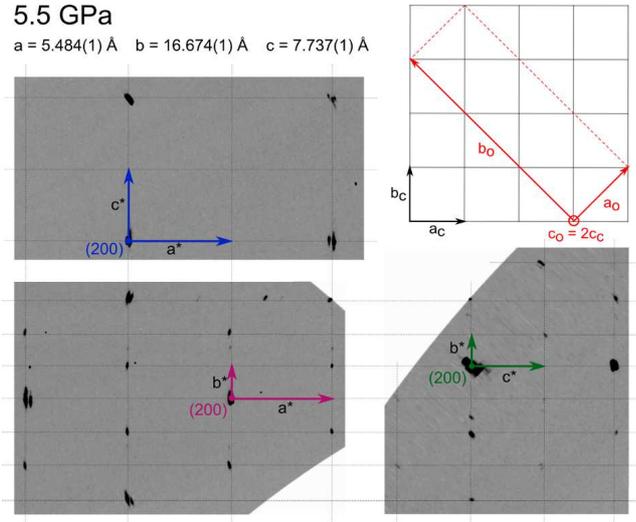}
\caption{Maps of $(hk0)$, $(2kl)$ and $(h0l)$ reciprocal planes obtained at 5.5 GPa. Due to the orientation of the sample in the DAC, we have almost no access to the $(0kl)$ plane and the sampling for $(2kl)$ is less precise. The relations of the orthorhombic unit cell axes to the simple cubic axes are recalled.}
\label{fig:phaseII}
\end{center}
\end{figure}

The diffraction patterns for all three phases can be indexed with an orthorhombic unit cell, taking into account the complex domain structure. This cell has lattice contants $a_o=5.484(1)$~\AA, $b_o=16.674(1)$~\AA, $c_o=7.737(1)$~\AA{} at 5.5~GPa and a comparatively large volume ($Z = 12$). For this cell, pseudo-cubic lattice constants can be calculated as $a_{\mathrm{pc}} = a_{\mathrm o}/\sqrt{2}$, $b_{\mathrm{pc}}=b_{\mathrm o}/(3\sqrt{2})$ and $c_{\mathrm{pc}} = c_{\mathrm o}/2$. In figure \ref{fig:phaseII}, we reproduce maps of reciprocal space at 5.5~GPa and give the relations between the lattice vectors of the orthorhombic cell and those of the parent cubic perovskite. 

Metrically, this cell is close to the orthorhombic cell proposed by Belik \etal{} \cite{Belik2009}, except that we find a $c$ axis doubled in all phases, as demonstrated at 5.5~GPa by the presence of the $(2k1)$ reflections in figure \ref{fig:phaseII}. The pattern can also be indexed with the $C2/m$ phase proposed by Haumont \etal{} \cite{Haumont2009}, but the diffraction does not give evidence for a monoclinic distortion. \modif{}{On the other hand, the orthorhombic cell identified by Kozlenko \etal{} in their neutron diffraction study \cite{Kozlenko2011}, with lattice constants $\sqrt{2}a_{\mathrm{pc}}\times 2\sqrt{2}a_{\mathrm{pc}}\times 2a_{\mathrm{pc}}$, cannot explain the observed patterns.}

The inspection of the reflections in figure \ref{fig:phaseII} shows the typical conditions for $I$ centering. In addition, the inspection of the $(h0l)$ plane shows that only reflections with $h,l=2n$ are present. This leaves the possibility for two extinction symbols $I-(ac)-$ and $I-cb$, which we cannot distinguish here because of an insufficient coverage of the $hk0$ plane due to the DAC geometry. This leads to 6 different possibilities among 4 space groups: $Ima2$ (46), $I2cm$ (46), $I2cb$ (45), $Imam$ (74), $Imcm$ (74), and $Imcb$ (72).

Within this choice of space group assignments, we can tentatively infer information from the evolution of the domain structure at the transition II$\longrightarrow$III. For a ferroic transition between two orthorhombic space groups, the formation of twins can only occur for a transition from point group $mmm$ for the parent phase to $mm2$ or $222$ \cite{Janovec2006}. The simplification of the domain structure therefore suggests that phase II has a space group $Ima2$, $I2cm$ or $I2cb$ while phase III has $Imam$, $Imcm$ or $Imcb$. For the transition III$\longrightarrow$IV, on the other hand, the diffraction patterns does not show any sign of a symmetry change, nor any change in the domain structure. This leaves the possibilities of a transition between any of the three centrosymmetric space groups mentioned above, which includes isostructural transitions. 

\subsection{High-pressure region: 11--60 GPa}

In the high-pressure region, we report two phase transitions at 38 and 48 GPa. Each phase will be separately described. Finally, we discuss the issue of non-hydrostatic stress conditions.

\subsubsection{Phase V: 11--38 GPa $(Pnma)$}

\begin{figure}[tb]
\begin{center}
\includegraphics[width=0.47\textwidth]{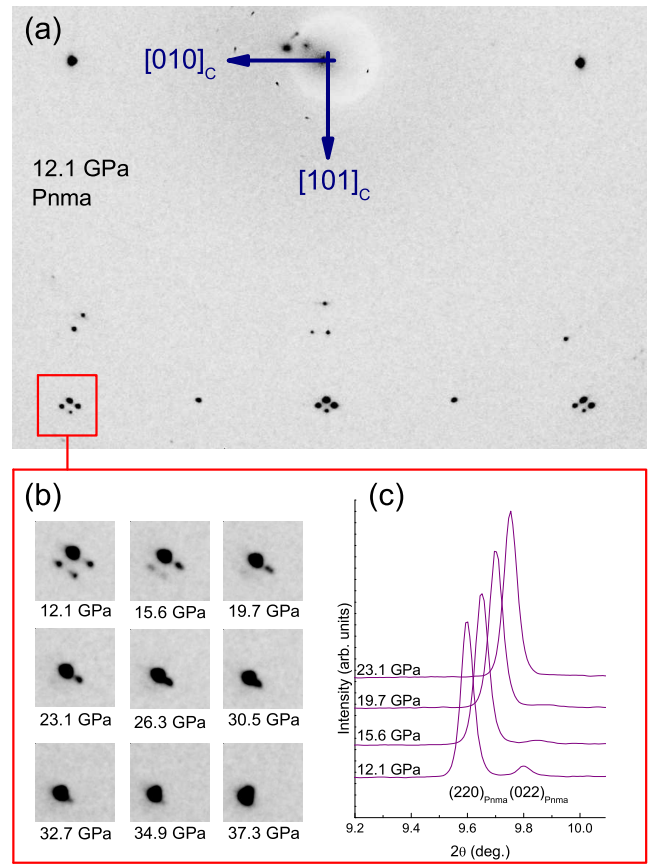}
\caption{(a) Diffraction pattern of a BiFeO$_3$ single crystal in the $Pnma$ phase integrated over the full angular range. The peaks show a splitting associated to the domain structure. (b) Evolution of the $(222)_C$ reflection under pressure and (c) selected corresponding $2\theta$ scans.}
\label{fig:DomainsPnma}
\end{center}
\end{figure}

Between 11 and 38 GPa, we find the now well established GdFeO$_3$-type $Pnma$ phase. A representative diffraction pattern is shown in figure \ref{fig:DomainsPnma}. In this phase, the crystal shows peak splitting associated with a four-variant domain structure, as is commonly observed for $Pnma$ crystals. As the pressure increases however, the intensity ratio changes and reflects the formation of a quasi single domain state (figure \ref{fig:DomainsPnma} (b) and (c)). 

\subsubsection{Phase VI: 38-48 GPa}

\begin{figure}[tb]
\begin{center}
\includegraphics[width=0.47\textwidth]{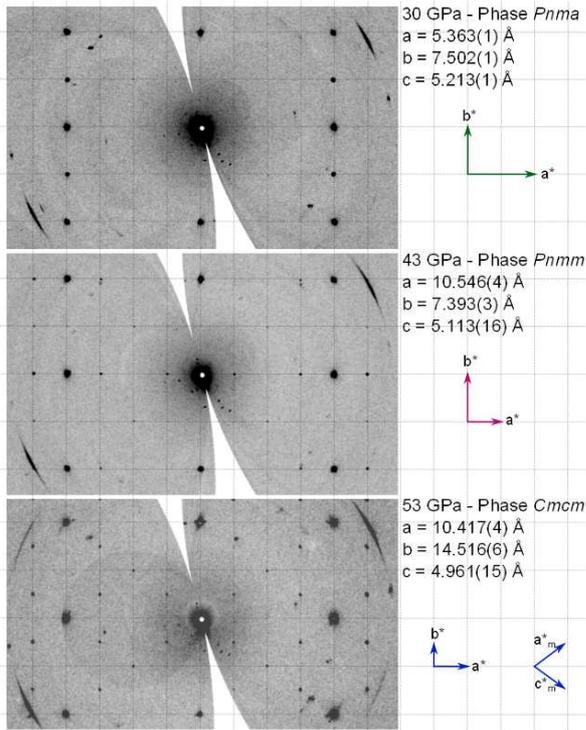}
\caption{Reciprocal space maps of the $(hk0)$ plane at 30 GPa (Phase $Pnma$), 43 GPa (Phase $Pnmm$) and 53 GPa (Phase $Cmcm$), taken from experiment 2 (Neon as PTM). The images have been rescaled for convenience so that the overall volume reduction is not apparent. The lattice constants and reciprocal lattice vectors for the different cells are recalled, including the monoclinic axes for the very last cell.}
\label{fig:phaseV}
\end{center}
\end{figure}

At 38~GPa, additional sets of superstructure appear as can be seen in figure \ref{fig:phaseV} on the reciprocal space map of the $(hk0)$ plane. Metrically the new cell can be derived from the $Pnma$ cell by a doubling of the $a$-axis. At 43~GPa, we find $a_{\mathrm o}=10.546(4)$~\AA, $b_{\mathrm o}=7.393(3)$~\AA{} and $c_{\mathrm o}=5.113(16)$~\AA{}. For this phase, we have $Z=8$ and we calculate pseudo-cubic lattice constants as $a_{\mathrm{pc}} = a_{\mathrm o}/(2\sqrt{2})$, $b_{\mathrm{pc}}=b_{\mathrm o}/2$ and $c_{\mathrm{pc}} = c_{\mathrm o}/\sqrt{2}$. Note that diffraction spots of the $(100)$ reflexions are clearly visible, but $(200)$ are not seen, which can only be explained by their weak intensity. Also, in this pattern, a small portion of a second crystal is visible, but is disoriented with respect to the main crystal, which may indicate that it comes from a piece of the crystal broken rather than a phase transition induced twin. The intensity of this additional peaks is however much weaker than the Bragg peaks of the main crystal and makes any confusion unlikely.

The analysis of the reflection conditions leads to two possible extinction symbols ($Pb\!-\!-$ and $Pn\!-\!-$). These two possibilities differ by a condition $0kl$:$k$ and $0kl$:$k+l$ respectively. Unfortunately, our collected single crystal pattern does not allow to distinguish the two possibilities because of the restricted coverage of the $0kl$ plane due to the experimental geometry and the orientation of the crystal in the cell. We found that the weak $(100)$ reflections, although unambiguous in the single crystal pattern, cannot be seen in the powder diffraction pattern. The ratio of the intensities of these weak superstructure to the main Bragg peaks is about 1/10000. Powder diffraction alone would fail to identify the correct unit cell and did not allow to discriminate between the two possibilities $Pb\!-\!-$ and $Pn\!-\!-$. We are left with 5 possible space groups: the 3 polar groups $Pb2_1m$ (26), $Pbm2$ (28) and $Pnm2_1$ (31) and 2 non polar groups $Pbmm$ (51) and $Pnmm$ (59). Based on the generally accepted fact that ferroelectricity vanishes under pressure and that there is no known ferroelectric perovskite beyond 15 GPa, we consider the last two possibilities as the most probable. In addition, we note that the $Pnmm$ opens the possibility for a group-subgroup relation with the $Pnma$ phase. 

\subsubsection{Phase VII: $P>48$ GPa}

The transition from phase VI to phase VII is evidenced by a clear change in the diffraction pattern. A reciprocal space map is presented in figure \ref{fig:phaseV}. This pattern can be indexed in a $C$-centered orthorhombic cell with the same axis directions as in phase VI. This new orthorhombic cell is derived from the $Pnmm$ cell by a doubling of the $b$ axis leading to lattice constants $a = 10.417(4)$~\AA, $b = 14.516(6)$~\AA{} and $c = 4.961(15)$~\AA{} at 53~GPa, with 16 formula unit per unit cell. With this indexing, the analysis of the reflection conditions leads to the extinction symbol $C\!\!-\!\!c-$ which allows for three different space groups: $Cmc2_1$ (36), $C2cm$ (40) and $Cmcm$ (63). The first two space groups are polar. With the same argument as before, we regard $Cmcm$ as the most probable space group for this phase.

Alternatively, the pattern can be indexed in a monoclinic cell. This cell has a volume half of the orthorhombic cell ($Z=8$) and reciprocal lattice vectors related to the orthorhombic by $\mathbf a_M^* = \mathbf a_O^* + \mathbf b_O^*$, $\mathbf c_M^* = \mathbf a_O^* - \mathbf b_O^*$, $\mathbf b_M^* = \mathbf c_O^*$, as depicted in figure \ref{fig:phaseV}. The orthorhombic and monoclinic cells are equivalent in the limit $\mathbf a_M^* = \mathbf c_M^*$. The analysis of the extinctions leave the possibilities for 5 monoclinic space groups for which no reflection condition applies: $P2$ (3), $P2_1$ (4), $Pm$ (6), $P2/m$ (10), $P2_1/m$ (11). Discarding as before the polar space groups, the two space groups $P2/m$ and $P2_1/m$ appear as more likely.

\subsubsection{Importance of non-hydrostatic stress}

As long as the pressure-transmitting medium (PTM) remains liquid ($P<12$~GPa for He, 4~GPa for Ne\modif{}{, 10~GPa for the classical 4:1 methanol--ethanol mixture}), the stress field can be safely regarded as hydrostatic. However, above this limit, a so-called deviatoric component may add to the hydrostatic stress field. The deviatoric stress has a rotational symmetry around the loading axis of the cell, with a stronger compressive stress along the axis, as checked by Zhao \etal{} \cite{Zhao2010} in an experiment with silicon oil as PTM. Non hydrostatic stress should then be regarded as unavoidable at high enough pressure, even though they are very often (legitimately) neglected when noble gases, particularly helium, are used as PTM. It is nonetheless of concern, for a meaningful comparison between different experimental works and with ab-initio calculations, to consider how such non-hydrostatic stress affects the outcome of an experiment. This can be done for example by comparing experiments performed in different PTM or, for single crystal investigations, by comparing experiments with crystals of different orientations.

In a previous diffraction work on BiFeO$_3$ single crystals in the 0-10~GPa range, we have shown that strongly non-hydrostatic stress have a remarkable effect on the phase transitions, by stabilizing a structure that is not observed in hydrostatic conditions \cite{Guennou2011}. Here, we have shown an evolution of the domain structure in the $Pnma$ phase (figure \ref{fig:DomainsPnma}), which can be explained by the onset of non-hydrostatic stress favoring one particular domain orientation. However, we also make the following observations. First, the same phases were observed in two experiments performed with different pressure-transmitting media (helium vs. neon). Moreoever, the powder diffraction experiment shows the same transitions, although with sligthly different pressures. Therefore, we believe that non-hydrostaticity of the stress field plays a minor role and is only likely to shift transition pressure and affect slightly the evolution of the lattice constants, but without affecting the structure of the phases.

\modif{}{In addition, complications may arise from the ferroelastic character of BFO. As the crystal undergoes different phase transitions, ferroelastic domain formation and reorientation may act as stress relief and modify the stress state of the sample. This can make hydrostatic conditions difficult to be achieved, even at low pressures. Ferroelasticity in BFO is unfortunately not extensively studied in literature, but deserves further attention in the light of the above comments.}

\subsection{Volume, lattice constants and strain}

\begin{figure}[tb]
\begin{center}
\includegraphics[width=0.47\textwidth]{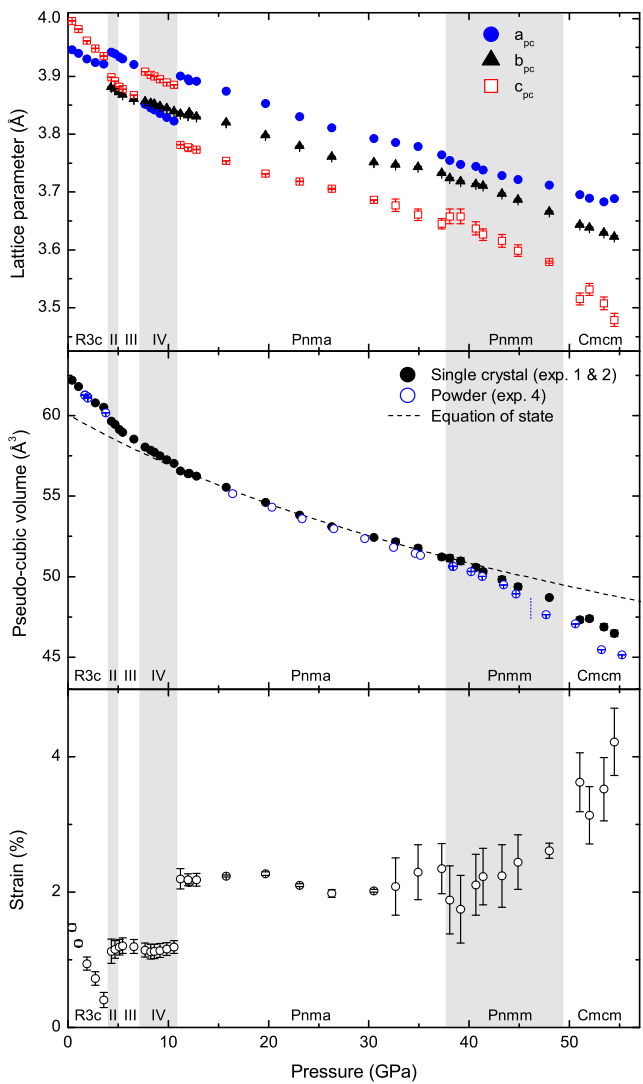}
\caption{Evolutions of the pseudo-cubic lattice constants (top), volume (middle) and total strain $e_{\mathrm{tot}}$ (bottom) in the full pressure range investigated. Data are from the single crystal experiments 1 and 2. In the volume plot, \modif{}{the data from the powder diffraction experiment have been added for comparison, and} the volume predicted by the equation of state has been extrapolated over the full pressure range as a guide to the eye.}
\label{fig:volumeandstrains}
\end{center}
\end{figure}

In figure \ref{fig:volumeandstrains}, we present the evolution of the lattice contants, volume and strain over the full pressure range. The strain is calculated as follows. As a reference state for all the phases, we choose a cubic cell with lattice constant $a_0=(V/Z)^{1/3}$. The general expressions for the strain components $e_{ij}$ can then be found in \cite{Salje1993}. In the orthorhombic system, the expressions reduce to $e_{11} = a_{\mathrm{pc}}/a_0-1$, $e_{22} = b_{\mathrm{pc}}/a_0-1$, $e_{33} = c_{\mathrm{pc}}/a_0-1$, where $a_{\mathrm{pc}}$, $b_{\mathrm{pc}}$ and $c_{\mathrm{pc}}$ are the pseudo-cubic lattice constants. We then calculate the "total" strain using Aizu's expression $e_{\mathrm{tot}}=\sqrt{\sum e_{ij}^2}$. This choice is somewhat arbitrary, but enables us to quantify and compare the strain in the different phases.

In the evolution of the lattice constants, the most remarkable feature is the transition III$\longrightarrow$IV that is marked by a "switching" of the $a_\mathrm{pc}$ and $c_\mathrm{pc}$ parameters for a volume (and a strain) that shows no significant jump, as already reported by Belik \etal{} in their powder diffraction study \cite{Belik2009}. We have no precise decription of the microscopic mechanism underlying this phase transition, but we note that a behavior of lattice constants similar to the present case was observed in BiMnO$_3$ \cite{Kozlenko2010} under pressure, where it was associated to a change in the magnetic order.

The evolution of the strain reveals the different distortion of the successive phases. With increasing pressure, the structure tends first toward a metrically cubic structure, whereby the strain approaches zero. This evolution is in good agreement with the literature \cite{Haumont2009}. The strain does not change appreciably through the transitions II$\longrightarrow$III$\longrightarrow$IV. Then, it shows a jump at the IV$\longrightarrow Pnma$, which appears more distorted than the original $R3c$ structure, and remains constant within experimental uncertainty through the $Pnma$ stability range. At higher pressures, above 40~GPa, the tendency of an increase of the distortion away from a cubic metricity is very clear.

In the volume plot, we have also plotted the values determined from the powder XRD experiment. The volume measured from the powder appears to deviate slightly from the values measured on the single crystal, with might be due to the different pressure-transmitting media used (helium vs. neon). A fit of the pseudo-cubic volume in the stability region of the $Pnma$ phase with a 2$^\mathrm{nd}$-order Birch-Murnaghan equation leads to $V_{\mathrm{12GPa}} = 56.43(8)$~\AA$^3$ and $K_{\mathrm{12GPa}} = 218(5)$~GPa for $K'_{\mathrm{12GPa}}$ being fixed to 4. This value compares reasonably well with the previously published values of 238 GPa \cite{Haumont2009}.

\modif{}{Last, we discuss the order of the transitions. From the discontinuities in the volume and lattice constants, it is clear that the transitions $R3c\rightarrow$II, III$\rightarrow$IV and IV$\rightarrow Pnma$ at 4, 7 and 11~GPa, respectively, are of first order. Transitions II$\rightarrow$III and $Pnma\rightarrow Pnmm$ at 5 and 38~GPa, on the other hand, present possible group--subgroup relationships and show no appreciable volume discontinuities, rather suggesting second-order transitions. This also seems to be the case for the last transition $Pnmm\rightarrow Cmcm$ transition at 48~GPa, although it is difficult to be conclusive, due to the uncertainty on the volume measurements and the limited number of points close to the transition.}


\section{Discussion}

\subsection{Multiple phases of BFO}

Perovskites are known for their richness of structural and physical phase transitions with different possible instabilities which can coexist and compete. Both, a fundamental and technological interest derive from this characteristic. One consequence of the various instabilities is an often complex phase diagram of perovskites, for instance in the $P$--$T$ or $x$--$T$ space which are the most explored diagrams. Besides perovskite-type manganites, ferroelectric perovskites are among the most investigated materials with the representative examples PbZr$_{1-x}$Ti$_x$O$_3$ (PZT) for the $x$--$T$ space \cite{LinesAndGlass,Noheda2002,Noheda2006} and BaTiO$_3$ (BTO) or KNbO$_3$ (KNO) for the $P$--$T$ space \cite{Iniguez2002,Pruzan2007}. Both examples display complex phase diagrams. Nevertheless, it is important to realize that when only one parameter ($P$, $T$ or $x$) is changed no more than 3 phase transitions are observed, e.g. the rhombohedral-orthorhombic-tetragonal-to-cubic phase sequence in BTO or KNO under pressure \cite{Iniguez2002,Pruzan2007}. Most other materials display fewer phase transitions under pressure: e.g. CaTiO$_3$ (no transition \cite{Guennou2010a}), SrTiO$_3$ (one transition \cite{Guennou2010}), or PbTiO$_3$ (two transitions \cite{Janolin2008}). In the light of this, the here observed 6 phase transitions of BiFeO$_3$ are remarkable. It is plausible to relate this observation to the fact that BFO presents at ambient conditions the rare coexistence of octahedra tilts, cation displacements and magnetic order.

A recent study by Di\'eguez \etal{} \cite{Dieguez2011} reinforces this intuitively expected intrinsic richness of structural phases in BFO by performing a systematic search for potentially stable phases by using first-principles methods. As a major outcome of their work, it has been demonstrated that BFO can present an unusual large number of (meta)stable structures of which the balance or stability is determined not only by the traditional soft modes but also by secondary modes. Our observation of a pressure-induced rich phase sequence corroborates experimentally the predicted structural richness of BFO. However, our experimental results go beyond the predictions in the sense that we report new phases with unusual large cell dimensions, illustrating an even more complex energy landscape than suggested by Diéguez's calculations \cite{Dieguez2011} which were restricted for computational reasons to smaller cells. The physical origin of such large unit cells is not yet understood, although its systematic occurrence in the low-pressure regime of Bi-based perovskites such as BiMnO$_3$, BiScO$_3$ and BFO has been hypothetically related to the presence of the so-called lone-pair in bismuth (see discussion in Ref. \cite{Haumont2009}). We note however, that we have shown here that BFO presents such large cells also at very high pressure, in sharp contrast to BiMnO$_3$ where the unit cells at very high-pressure are small \cite{tobepublished}. This suggests that the mechanism for very large unit cells has to be understood beyond the simple presence of Bi and thus deserve further attention, namely through adapted first-principle calculations. NaNbO$_3$ \cite{Mishra2011} and AgNbO$_3$ \cite{Fu2007} show similarly large unit cells that have been interpreted in terms of complex tilt systems. Here, we understand by "complex tilt systems" collective tilts of the oxygen octahedra, whose description requires more than a doubling of the lattice constants of the cubic primitive perovskite cell and as such go beyond the now classical approach by Glazer \cite{Glazer1972,Glazer1975}. These compounds also show a coexistence between octahedra tilts and (anti)ferroelectric cation ordering and might provide some understanding. 

\subsection{Stability and compressibility of the $Pnma$ phase}

A particular phase within this complex structural landscape is the orthorhombic $Pnma$ phase. Generally speaking, $Pnma$ is one of the very common structures among perovskites and is known to withstand very high pressures. In particular, the rare-earth orthoferrites $RE$FeO$_3$ maintain their $Pnma$ structure at high-pressure, even through their large volume collapse associated with a spin transition around 50 GPa \cite{Adams2009}. In BFO, it presents a remarkable stability from 11~GPa up to 38~GPa. This constrasts with the effect of high temperature where it persists only in a narrow temperature range \cite{Arnold2010}. The stability is in agreement with calculations \cite{Dieguez2011} that have predicted that the $Pnma$ phase constitutes BFO's most stable phase beside the room temperature $R3c$ phase.  

In the $Pnma$ phase, the evolution of the tilt angles under pressure is of particular interest due to the importance of these angles for the electronic band structure. The two tilt angles Ti-O1-Ti ($\alpha_1$) and Ti-O2-Ti ($\alpha_2$) that characterize the $Pnma$ structure \cite{Mitchell2002} are best determined from the atomic positions obtained by Rietveld refinements, but the quality of our powder data does not allow to determine atomic position with satisfying uncertainties. Tilt angles can also be calculated from the lattice constants under the assumption of undistorted polyhedra \cite{Mitchell2002}, but these formulas are known to give only rough estimations of the tilt angles \cite{Guennou2010a}. From these formula and our measured lattice constants, we can estimate that the octahedra tilt angles decrease in the $Pnma$ structure by less than 2$^\circ$ between 10 to 38~GPa, with only little changes in the spontaneous deformation (figure \ref{fig:volumeandstrains})

Following Angel \etal{} compressibility rules \cite{Angel2005} the tilt angle is predicted to decrease with increasing pressure for most $A^{3+}B^{3+}$O$_3$ perovskites, while it increases for $A^{2+}B^{4+}$O$_3$ perovskites. Our experimental observation appears in qualitative agreement with this rule, in spite of the presence of the lone pair of bismuth. Such small changes have been reported \cite{Zhao2004} to occur in orthoferrites in the presence of large and heavy $A$ cations, as is the case here for BFO. In future works, it would be interesting to investigate in more detail rare-earth orthoferrites in a similar manner to obtain a better understanding of the particular role of bismuth on the compressibility of the perovskite.

\subsection{Implications for the insulator-to-metal transition}

Another aspect of the very high-pressure regime is the reported occurrence of an insulator-to-metal (MI) phase transition accompanied by the disappearance of magnetic order around 50 GPa \cite{Gavriliuk2005,Gavriliuk2006,Gavriliuk2007,Gavriliuk2008}. There is an ongoing discussion in the literature on the origin of this MI transition in terms of a high-spin to low-spin crossover, a Mott-type transition and related changes in crystal symmetry or not \cite{Gavriliuk2008,Catalan2009,Gonzalez-Vazquez2009,Rozenberg2005,Lyubutin2009}. Our structural study does naturally not allow discussing in detail the proposed models of this physical phase transitions. However, our study provides evidence that the MI transition is symmetry breaking, and it rules out a transition to the earlier considered cubic \cite{Catalan2009} or rhombohedral \cite{Gavriliuk2008,Gonzalez-Vazquez2009} phases. Rare-earth orthoferrites $RE$FeO$_3$ present similar transition around 50 GPa which has been assigned to a low-to-high spin phase transition of the Fe$^{3+}$ cations accompanied by a steep anomaly in the cell volume, particularly marked in LuFeO$_3$ \cite{Adams2009,Pasternak2002}. Such a steep anomaly, characteristic for low-to-high spin phase transition is not observed in our measurements, thus rather questioning such a scenario.

\section{Concluding remarks}

\begin{figure}[tb]
\begin{center}
\includegraphics[width=0.47\textwidth]{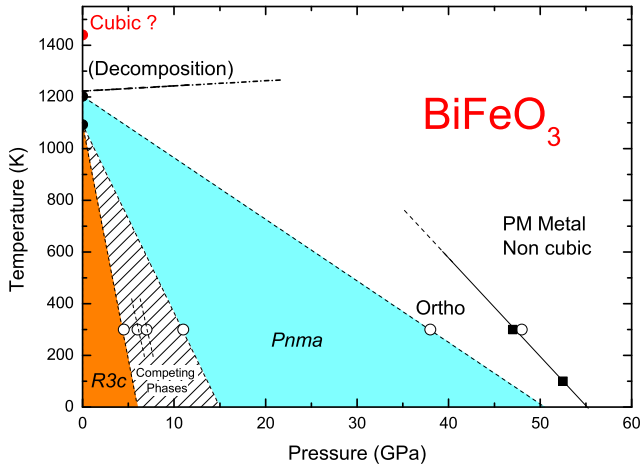}
\caption{Revised schematic pressure--temperature phase diagram of BFO. The symbols are experimental points taken from this work \modif{}{(open symbols)} and Refs. \cite{Catalan2009,Arnold2010} \modif{}{(full symbols)}. The transition temperature to a cubic phase at ambient pressure lies beyond the decomposition temperature and is a theoretical prediction \cite{Kornev2007a}.}
\label{fig:phasediagram}
\end{center}
\end{figure}

In this study we have presented a structural investigation of the model multiferroic perovskite BiFeO$_3$ under high pressure up to 60~GPa. The complementary use of synchrotron X-ray diffraction and Raman scattering reveals a remarkable richness of 6 phase transitions in the investigated pressure range. The occurrence of these six pressure-induced phase transitions together with the four transitions as a function of temperature reported in the literature \cite{Catalan2009,Palai2008,Haumont2008a,Dieguez2011} suggests a very complex phase $P$--$T$ phase diagram that is characterized by the competition of the BFO-characteristic instabilities: octahedra tilts, cation displacements, magnetic order and MI phase transitions.

In a first attempt, Catalan and Scott \cite{Catalan2009} have suggested in 2009 a preliminary and very useful phase diagram. Since then, our understanding of the different phase transitions has increased and we propose in Fig. \ref{fig:phasediagram} an updated schematic phase diagram, which is based on both literature results and the outcome of our present pressure work. In the following, we will discuss the different features of this phase diagram with a particular emphasis on the essential points of our present work.

\begin{itemize}
\item[(i)] In agreement with the calculations by Di\'eguez \etal{} \cite{Dieguez2011}, the phase diagram is characterized by two particularly stable phases. First, the rhombohedral $R3c$ phase (orange in Fig. \ref{fig:phasediagram}) which is at ambient pressure stable over 1000~K. Despite this temperature stability, the $R3c$ structure is rapidly destabilized under hydrostatic pressure and disappears at a modest pressure of 4~GPa. Second, the orthorhombic $Pnma$ structure (blue), which is stable at 300~K between 11 and 39~GPa. We have shown that the $Pnma$ phase is characterized by only small changes in the spontaneous deformation (and octahedra tilts), so that the compression mainly acts through bond shortening.
\item[(ii)]	Between the stable $R3c$ and $Pnma$ phases, we observe under high-pressure three different phases of orthorhombic symmetry, thus a rather complex energy landscape within only a few GPa. All three intermediate phases are characterized by unusual large unit cells, implying complex tilt systems and possibly off-center cation displacements triggered by the lone pair of bismuth. In the phase diagram we consider this intermediate region (striped in Fig. \ref{fig:phasediagram}) as only one single region that is characterized by a multitude of competing phases, which might well reveal even more richness when varying the parameter temperature. \modif{}{The suggestion of a distinct antipolar phase by Kozlenko \etal{} \cite{Kozlenko2011} supports this view.} We also remind that \modif{it has been shown that}{} new phases can easily be induced by non-hydrostatic conditions in this region \cite{Guennou2011}. The phase diagram is consistent with the idea of a complex deformation-temperature phase diagram as can be also explored by temperature-dependent investigation of strained thin films (although the biaxial strain in thin films can of course not be directly compared to hydrostatic conditions). The very recent reports \cite{Kreisel2011,Siemons2011,MacDougall2011,Infante2011} of a structural phase transition close to room temperature in highly compressively strained BFO thin films adds further support to this.
\item[(iii)] The very high-pressure regime is characterised by two further phase transitions at 38 and 48~GPa, which are again characterized by large unit cells. Interestingly, we observe in this regime an increase of the distortion away from the cubic parent perovskite cell, which contrasts the earlier sketched \cite{Catalan2009} high-pressure tendency towards a cubic structure. On the other hand, this is consistent with ab-initio calculations by Gonz\'alez-V\'azquez and \'I\~niguez according to which the simple cubic phase does not become the ground state even at high pressure \cite{Gonzalez-Vazquez2009}. At high temperature, the most recent studies \cite{Arnold2010} show that the cubic phase can only be expected at temperatures higher than the decomposition point and is therefore regarded as unstable. 
\item[(iv)] The insulator-to-metal phase transition in BFO is symmetry breaking. However, contrary to the earlier phase diagram \cite{Catalan2009} the pressure-induced paramagnetic metallic phase is not cubic, but rather of orthorhombic symmetry.
\end{itemize}

The above discussion and the phase diagram in Fig. \ref{fig:phasediagram} is based only on temperature measurement at 1~bar and pressure measurements at 300~K. As a consequence, the unexplored large intermediate $P$--$T$ region remains hypothetic and is likely to give rise to new and unexpected features. \modif{Based on our here presented work we encourage future work, particularly into the mid-pressure region with competing phases, the understanding of the reported unusual large unit cells, the mechanism at the MI phase transitions and, more generally, the exploration of the strain-temperature phase diagram, be it by hydrostatic pressure or strain in thin films.}{We strongly encourage the exploration of the strain-temperature phase diagram, be it by hydrostatic pressure or strain in thin films, and particularly into the mid-pressure region with competing phases.}

\section{Aknowledgments}

We are grateful to the ESRF staff, especially M. Mezouar for allocation of inhouse beamtime, D. Chernyshov and A. Bossak for their help with the softwares, and W. Crichton for fruitful discussions. Financial support from the French National Research Agency (ANR Blanc PROPER) is acknowledged.

\bibliographystyle{aip}
\bibliography{biblio}

\end{document}